\documentclass[journal]{IEEEtran}
\ifCLASSINFOpdf
\usepackage[pdftex]{graphicx}
\graphicspath{figures/}
\DeclareGraphicsExtensions{.pdf,.jpeg,.png}
\else
\usepackage[dvips]{graphicx}
\graphicspath{{figures/}}
\DeclareGraphicsExtensions{.eps}
\fi
\usepackage{color, soul}
\usepackage[cmex10]{amsmath}
\usepackage{bm}
\usepackage{hyperref}
\usepackage{mathtools}
\usepackage{helvet}
\usepackage{xcolor}
\usepackage{cite}
\usepackage{amsfonts}
\usepackage{multirow}
\usepackage{footnote}
\makesavenoteenv{tabular}
\usepackage{threeparttable}

\usepackage{url}
\begin{document}
\title{Variation-Resilient FeFET-Based In-Memory Computing Leveraging Probabilistic Deep Learning}

\author{Bibhas~Manna, Arnob~Saha, Zhouhang~Jiang, Kai~Ni,
and~Abhronil~Sengupta,~\IEEEmembership{Senior Member,~IEEE}% <-this % stops a space
\thanks{B. Manna, A. Saha, and A. Sengupta are with the School of Electrical Engineering and Computer Science, The Pennsylvania State University, University Park, PA 16802, USA. E-mail: sengupta@psu.edu.\\
\indent Z. Jiang and K. Ni are with the Department of Electrical Engineering, University of Notre Dame, IN 46556 USA. }}
\maketitle
\begin{abstract}
\small{Reliability issues stemming from device level non-idealities of non-volatile emerging technologies like ferroelectric field-effect transistors (FeFET), especially at scaled dimensions, cause substantial degradation in the accuracy of In-Memory crossbar-based AI systems. In this work, we present a variation-aware design technique to characterize the device level variations and to mitigate their impact on hardware accuracy employing a Bayesian Neural Network (BNN) approach. An effective conductance variation model is derived from the experimental measurements of cycle-to-cycle (C2C) and device-to-device (D2D) variations performed on FeFET devices fabricated using 28 nm high-$k$ metal gate technology. The variations were found to be a function of different conductance states within the given programming range, which sharply contrasts earlier efforts where a fixed variation dispersion was considered for all conductance values. Such variation characteristics formulated for three different device sizes at different read voltages were provided as prior variation information to the BNN to yield a more exact and reliable inference. Near-ideal accuracy for shallow networks (MLP5 and LeNet models) on the MNIST dataset and limited accuracy decline by $\sim$3.8-16.1\% for deeper AlexNet models on CIFAR10 dataset under a wide range of variations corresponding to different device sizes and read voltages, demonstrates the efficacy of our proposed device-algorithm co-design technique.}
\end{abstract}

\begin{IEEEkeywords}
Variation-Aware Design, FeFET Crossbar, Device-Algorithm Co-Design, Bayesian Neural Network, In-Memory Computing.
\end{IEEEkeywords}

\section{Introduction}
Emerging non-volatile memories capable of performing simultaneous compute and storage functionalities show great promise for the hardware acceleration of deep neural networks \cite{b1, b2}. The data-intensive and complex vector-matrix multiplication operations required in neural networks can be realized on-chip by harnessing the inherent physical attributes of the memory devices arranged in an array fashion - resulting in ``In-Memory Computing". Among different potential memory candidates such as resistive random-access memory (RRAM), phase-change memory (PCM), magnetic devices, etc., hafnia-based FeFET has lately earned great interest due to its CMOS compatibility, low energy operation, multilevel programming capability with wider dynamic range, decoupled read-write operation, easy array-level integration, among others \cite{b3, b4, b5, b6}. The voltage-driven partial polarization switching in the ferroelectric layer of FeFET promotes gradual tuning of channel conductivity, mimicking analog synaptic weight update behavior. However, process variation induced stochastic variabilities stemming primarily from the poly-crystalline ferroelectric and their pronounced effect with device scaling poses a serious challenge to accomplishing reliable computing using FeFET crossbars. The device-level non-idealities with read-write fluctuations cause the stored weight (i.e., programmed conductance) to deviate significantly from the expected trained value, resulting in drastic accuracy degradation of the neural network at the hardware level. Thus, addressing device-level reliability and proposing practical solutions to combat their consequences are crucial to designing variation-tolerant FeFET based neuromorphic computing. 

Prior efforts in this direction mostly adopt either expensive retraining or repeated evaluation-remapping methods demanding non-trivial design overhead \cite{b7, b8, b9}. Some works incorporate generalized noise models in the network weights at the algorithm level and attempt to compensate for their effects through iterative training but are unable to perform the learning task jointly with robustness optimization \cite{b10, b11}. Bayesian inference-based approach on memristor-crossbar based systems considering device non-idealities and stuck-at-faults has been proposed recently to achieve robust computing  \cite{b12, b13}. However, the proposal is formulated based on a parameterized canonical form of variation derived from a more generalized and hypothetical device model and therefore does not reflect realistic interplay of variations with device dimensions and operating voltage conditions. This work seeks to bridge this critical gap by underscoring the strong need to consider such hardware-software co-design effects in algorithm design supported by extensive experimental variation characterization on industry-scale FeFETs and subsequent performance evaluation on standard machine learning benchmark suites. 
%In this work, we propose variation-aware design of FeFET based In-Memory AI hardware accelerators adopting a unique device-algorithm co-design approach.
An empirical conductance variation model derived at the individual device level is coupled to the probabilistic learning based uncertainty optimizer at the algorithm level to abate the effect of hardware non-idealities on recognition performance. The work possesses the following distinctive novelties:
\begin{itemize}
    \item Formulation of a realistic conductance variation model capturing relationship of device non-idealities with scaling and operating voltage conditions through direct experimental characterization of scaled FeFET devices fabricated on industry standard process technology.
    \item Design of a practically feasible variation-aware framework by incorporating device size-dependent unique variation characteristics against various programming voltages into the BNN training framework to deliver robust and stable inference. 
\end{itemize} 

%The rest of the paper is arranged as follows: Section II explains the scheme for experimentally characterizing the channel conductance of FeFETs having different gate dimensions. Section III provides preliminaries on the operation of BNN and how it accomplishes robustness and accuracy maximization tasks simultaneously. Section IV derives an effective size-dependent device non-ideality model, formulates the prior for BNN, and discusses the efficacy of the proposed device-algorithm co-design framework to mitigate the impact of device-level non-idealities on computing performance. Finally, we outline the key conclusion of our work in section V. 

The rest of the paper is organized as follows: Section II explains the scheme for experimentally characterizing the channel conductance of FeFETs having different gate dimensions. Section III presents design of the proposed Bayesian training framework incorporating device level conductance variation information. The section provides necessary background on the operation of BNN, explores the impact of device scaling on channel conductance properties through simulations and experiments, derives an effective device-specific non-ideality model, and formulates the prior for BNN. Section IV evaluates performance of the proposed variation-aware framework and discusses its efficacy to mitigate the impact of device-level non-idealities on computing performance. Finally, we outline the key conclusions of our work in section V.  

\section{Experimental Characterization of FeFET}
% =======
% FIG. 01
% =======
\begin{figure} 
  \begin{center}
  \includegraphics[width=3.4in]{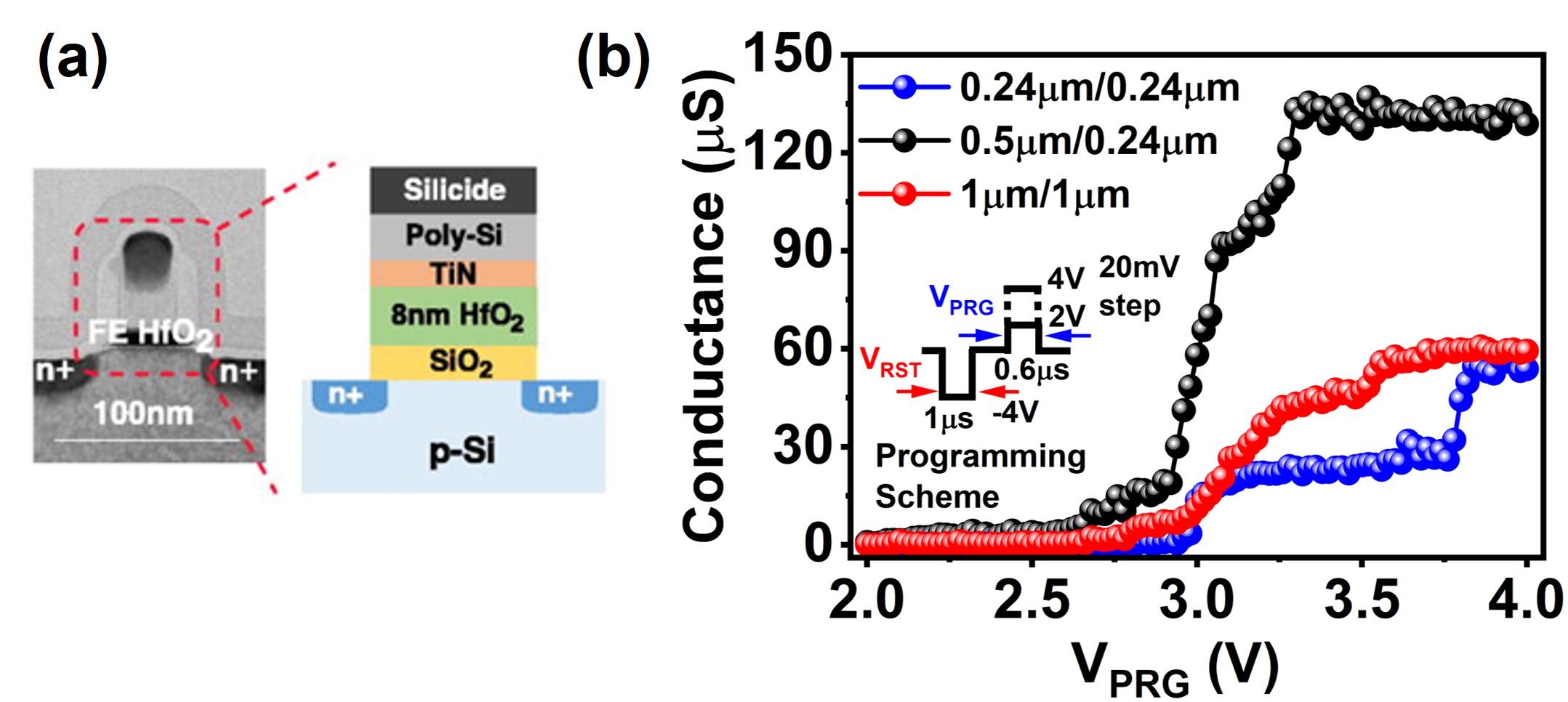}\\
  \caption{(a) TEM cross-section and schematic representation of FeFET fabricated on 28nm HKMG node with doped $HfO_{2}$ serving as the ferroelectric. (b) Conductance-programming voltage characteristics of FeFET for three different device dimensions at read voltage, $V_{Read}$, of 1.2V.}\label{Device_Schematic}
  \end{center}
\end{figure}
Experimental measurements have been performed on FeFET fabricated using an industrial 28nm high-$k$ metal gate (HKMG) technology node \cite{b14}. Three identical FeFET devices of different gate sizes: $W$(width)/$L$(length) = 1$\mu$m/1$\mu$m, 0.5$\mu$m/0.24$\mu$m, and 0.24$\mu$m/0.24$\mu$m  have been considered for the present study. The schematic device structure and corresponding cross-sectional TEM image are shown in Fig. 1(a)\cite{b14, b32}. Each device comprises a vertical metal-ferroelectric-insulator-silicon (MFIS) stack with 8nm thick doped $HfO_{2}$ and $\sim$1nm thick high-quality $SiO_{2}$ functioning as ferroelectric (FE) and insulator layers respectively. To measure the modulation of channel conductivity in response to polarization switching, each device was subjected to a gate voltage with a pulsing scheme, as illustrated in Fig. 1(b). A reset pulse ($V_{RST}$) of -4V preceded every programming pulse ($V_{PRG}$) to switch all the domains to the negative polarization state, thus resetting the device to the initial lowest conductance state every time. A positive programming pulse of progressively growing amplitude (2V to 4V with a step of 20mV) was employed to access all possible intermediate conductance states of the device. The read-out of the programmed state was accomplished immediately after each write operation by applying a ramp gate voltage ($V_{Read}$). The drain terminal was held at 0V during the reset and programming operations, and was switched to 50mV during the read operation. 

\section{Design of Variation-Robust Bayesian Training Framework}

\subsection{Preliminaries}
The proposed algorithmic framework leverages the intrinsic property of Bayesian neural networks to produce accurate and robust inference under weight fluctuations by incorporating the probability distributions associated with variation data obtained through extensive characterization of the FeFET devices considering effects of device sizing, read noise, among others. The $j^{th}$ weight, $w_{j}$, of a network is assumed to be mapped to the conductance state, $g_{j}$, within the FeFET programming range present at the $j^{th}$ cross-point, following the relationship \cite{b15}
\begin{equation}\label{weight_mapping}
    g_{j} = (\frac{g_{mx}-g_{mn}}{w_{mx}-w_{mn}})[{|}{w_{j}{|}}-w_{mx}]+g_{mx}
\end{equation}
where, $g_{mx}$, $g_{mn}$, and $w_{mx}$, $w_{mn}$ are the maximum and minimum values of the conductance and corresponding weight, respectively. As the programmed conductance is variational, it is highly appropriate to treat each network weight as a distribution rather than having a specific value, which is fortunately the inherent behavior of BNNs. For a given dataset, \textit{D}, the BNN accepts a priori, $P(w)$, on variation characteristics of noisy weights to find posterior weight distributions following the Bayes rule: $P(w{\mid}D)$ = $P(w)P(D{\mid}w)/P(D)$. As the true weight posterior, $P(w{\mid}D)$, is  computationally intractable, stochastic variational inference scheme is applied to approximate $P(w{\mid}D)$ with a distribution, $q(w{\mid}{\theta})$, that minimizes  the Kullback-Leibler (KL) divergence with true Bayesian posterior \cite{b16A,b16B}. The corresponding objective function, $\eta(D,\theta)$, for optimization becomes:
\begin{equation}\label{weight_mapping}
    \eta(D, \theta) = -E_{q(w{|}\theta)}[\log P(D{|}w)] + KL[q(w{|}\theta) {||} P(w)]
\end{equation}
The approximated posterior distribution of weights, $q(w{\mid}{\theta})$, in BNNs is learnt iteratively through the “Bayes by Backprop” method to enforce that the posterior follows the device variation characteristics supplied as prior information, $P(w)$, to the framework \cite{b16}. %For a given dataset, \textit{D}, 
Considering Gaussian distribution, variational parameter, $\theta(\mu_{q}, \sigma_{q})$, for each weight posterior of the BNN is updated by descending along the gradients of the objective function. A re-parameterization trick is useful to obtain more efficient gradient estimation for $\theta$, enabling training iteration to be compatible with standard backpropagation \cite{b16}. The data-dependent first component of Eqn. (2) represents likelihood cost, which is the standard loss function averaged over multiple single network models derived by sampling the posterior weight distribution. On the other hand, the prior-dependent second component represents the KL divergence loss, which computes the degree of dissimilarity between the prior and posterior distributions and accounts for ensuring robustness to the optimization problem. Hence, the framework is capable of simultaneously handling the goal of maximizing accuracy and minimizing  reliability induced errors by driving the posterior to follow the prior through the backpropagation method. Upon successful completion of training, the mean values, $\mu_{q}$, of the optimized posterior distributions are regarded as the fully trained weights (mapped to the FeFET conductance at the hardware level) for inference evaluation.
\subsection{Conductance Characteristics of Scaled FeFET}
We started our analysis by measuring the conductance-programming voltage characteristics of FeFET for three different gate areas, as presented in Fig. 1(b). A more gradual transition of channel conductance with $V_{PRG}$ was observed for the $W/L$=1$\mu$m/1$\mu$m device. The gradual switching with a continuum of states reflects a broader distribution of coercive voltages across a large number of domains (tiny switchable units) in the FE layer such that polarization flipping is possible for a subset of domains at almost every incremental $V_{PRG}$. However, as the gate area shrinks, the number of domains in the FE layer reduces proportionally, and the non-homogeneity and randomness to the coercive field distribution becomes more pronounced. This certainly introduces non-linearity to the conductance programming profile with reduction in number of states, as evident from Fig. 1(b) for the $W/L$=0.5$\mu$m/0.24$\mu$m and 0.24$\mu$m/0.24$\mu$m device sizes. %The presence of an adequate number of distinctive conductance states between two extreme saturation and a reasonably high conductance on/off ratio imply the aptness of all three devices to implement analog synaptic functionalities.
% =======
% FIG. 02
% =======
\begin{figure} 
  \begin{center}
  \includegraphics[width=3.5in]{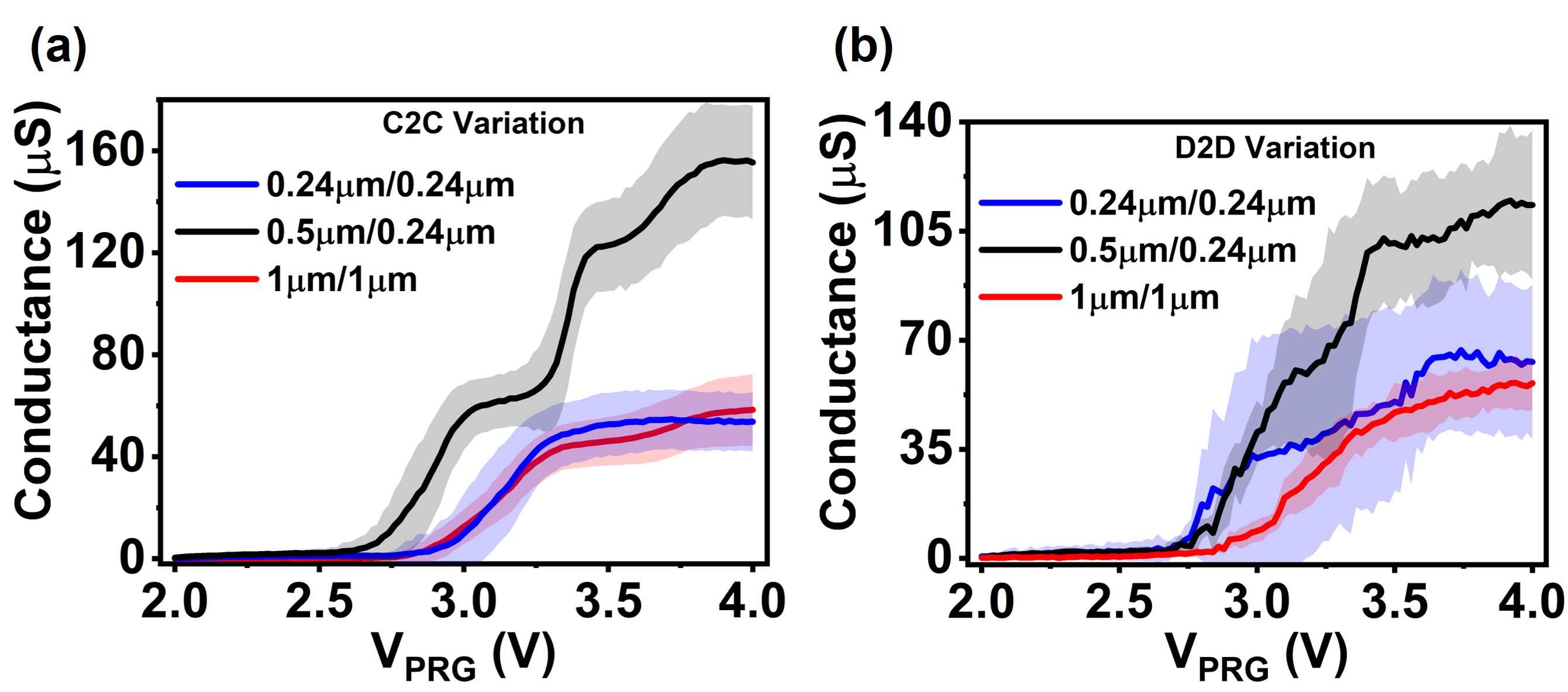}\\
  %\caption{(a) Cycle-to-cycle (C2C) variations measured for 50 consecutive programming cycles for each amplitude of $V_{PRG}$ corresponding to a single device of each device size and (b) Device-to-device (D2D) variations recorded over 3 devices of the same size by running a single programming pulse for each amplitude of $V_{PRG}$ at $V_{Read}$ of 1.2V. The solid line and corresponding broadening in each plot represent the mean and standard deviation of conductance variation respectively.}
  \caption{Filled error plot showing mean (solid line) and associated standard deviation (broadening) of conductance states for (a) cycle-to-cycle (C2C) variations measured over 50 consecutive programming cycles for each amplitude of $V_{PRG}$ corresponding to a single device of each device size and (b) device-to-device (D2D) variations recorded over 3 devices of the same size by running a single programming pulse for each amplitude of $V_{PRG}$ at $V_{Read}$ of 1.2V.}
  \label{Variation_data}
  \end{center}
\end{figure}

%\vspace{-2mm}
The stochastic polarization switching of individual domains in the FE layer and the process variations involved in device fabrication introduce obvious C2C and D2D variation effects on the conductance, especially for scaled devices. Fig. 2(a) and (b) illustrate the mean and standard deviation of experimentally measured C2C and D2D variations as filled error plots at a $V_{Read}$ of 1.2V. The intra-device variations measured over 50 cycles indicates that an appreciable amount of C2C variation is present for all devices and is almost insensitive to the device size. On the other hand, the inter-device deviations calculated over three devices for each size implies that D2D variations increases drastically with  device scaling and dominates over C2C variation for highly scaled devices. %It is worth mentioning here that we had to limit our experimental characterization to three devices for each size since we have to perform C2C measurements for each device individually in order to construct the cohesive variation model to be used in the algorithmic framework. 
The observation agrees well with earlier reports on scaled FeFETs \cite{b17, b18}. The higher D2D variation in smaller devices is primarily attributed to reduced domain number, increased in-homogeneity in the domain distributions, more randomness in the distribution of ferroelectric and dielectric phases in the FE layer, among others. \cite{b5, b19, b20, b21, b22}. Although the degree of D2D variations may differ if the experimental characterization is performed over larger number of devices, their dependency on device scaling is expected to remain unchanged.

Furthermore, we substantiated our observation on D2D variations computed over limited experimental data using a well-established Monte Carlo based simulation model \cite{b23, b24}. The model considers the poly-crystalline FE layer as an ensemble of multiple uncorrelated domains randomly initialized to either of the two stable polarization states. The switching between states for a domain at any time step, $\Delta t$, is associated with a finite probability, $P_{SW,i}$, which under the influence of temporally varying electric field, $E_{FE}(t)$, can be expressed as:
\begin{equation}\label{switching_probability}
    P_{SW,i} = 1-exp [h_{i}(t)^\beta -h_{i}(t+\Delta t)^\beta ]
\end{equation}
where, $\beta$ is the shape parameter of the probability distribution. The history parameter, $h_{i}(t)$, which is responsible for accumulating instantaneous stimuli to the $i^{th}$ domain over time can be computed as:
\begin{equation}\label{history_parameter}
    h_{i}(t) = \int_{t_{0}}^{t} \frac{dt^{'}}{\tau_{SW,i}(E_{FE}(t), E_{a,i})}
\end{equation}

\begin{figure} [t]
  \begin{center}
  \includegraphics[width=3.4in]{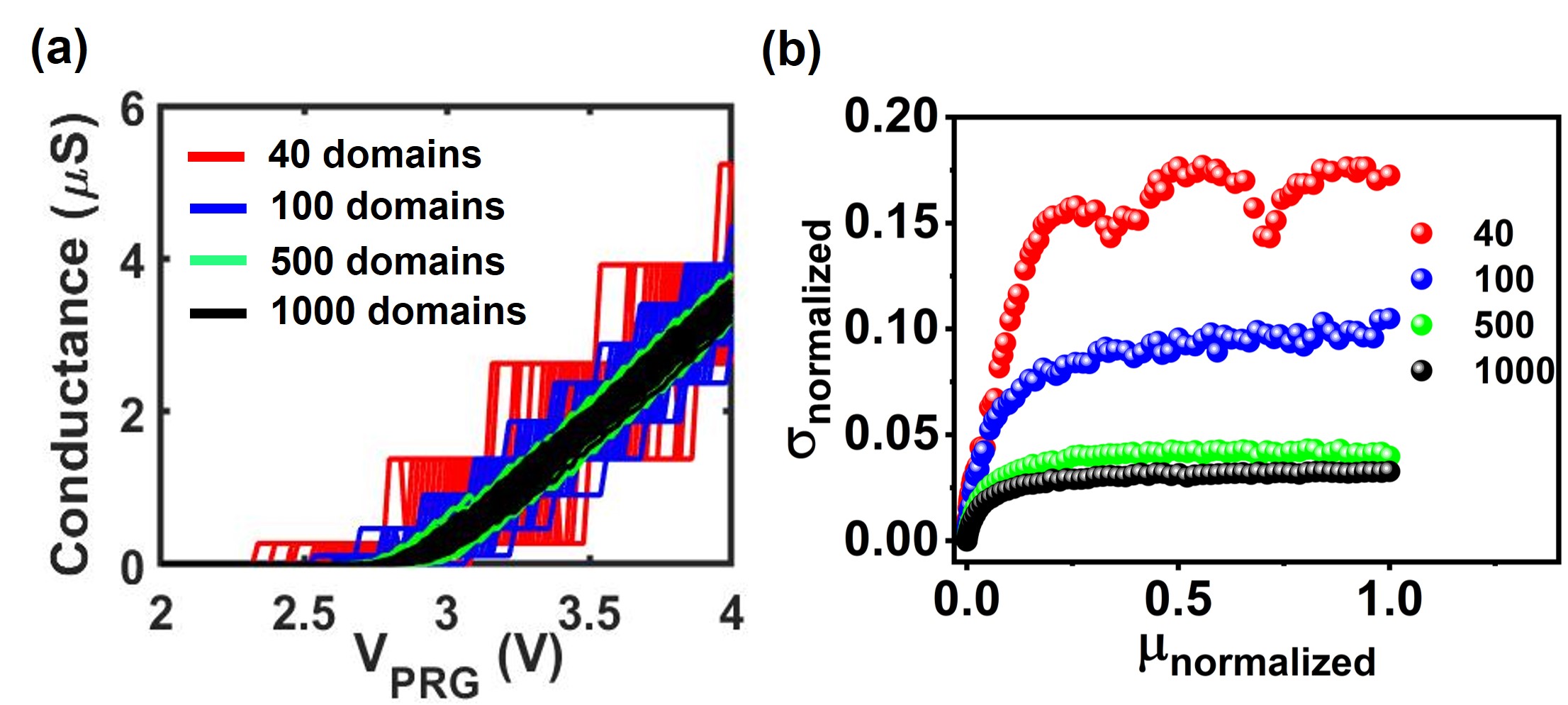}\\
  \caption{(a) Simulation results showing D2D variations computed over 200 devices for different number of domains in the FE layer. (b) The standard deviation plotted against mean of the D2D variations after normalizing the conductance programming data to a maximum value of unity.}
  \label{MC_simulation}
  \end{center}
\end{figure}

The domain switching time constant, $\tau_{SW,i}$, can be formulated following the nucleation limited switching model \cite{b23}. The parameter, $h_{i}(t)$, captures polarization accumulation effects and increases over time until the domain reverses its state. The polarization of the entire film at any time is estimated as a summation over all the individual domains. The time-dependent polarization dynamics of the film is next coupled to the conventional charge-voltage equation of the n-channel FET to solve for the channel conductance of the FeFET self-consistently. The values of the parameters used for the device simulation are the same as noted in prior work \cite{b25, b26}. For the simulation of D2D variations, an activation field value is sampled randomly over a normal distribution of $E_{a}$ for each of the domains in 200 identical devices \cite{b18}. The activation field affects the switching probability, $P_{SW,i}$,  of the individual domains through the history parameter, $h_{i}(t)$ (in Eqn. (4)). Since the distributions of $E_{a}$ are not identical across 200 devices containing the same number of domains, the partial polarization switching dynamics of the FE layer is expected to be different from one device to another. The simulated D2D variation, as shown in Fig. 3(a), demonstrates that the variation increases greatly with the decrease in domain number (i.e., with the down-scaling of the device area), thereby in agreement with our experimental findings.  

\subsection{Device Variability Modeling}
To quantify the amount of variation involved in the  conductance programming process and model its dependence on  device sizing, an effective variation parameter combining both the spatial (D2D) and temporal (C2C) effects was derived based on experimental data. The mean, $\mu_{com}$, and standard deviation, $\sigma_{com}$, of the combined variation effects was estimated by averaging  C2C variations over multiple devices. Fig 4(a) illustrates such a combined variation profile within the entire programming range for different device sizes at two read voltages of 0.6V and 1.2V. The exponential rise of $\sigma_{com}$ for initial smaller values of $\mu_{com}$ is primarily due to the greater degree of randomness associated with bias-dependent domain polarization switching at relatively weaker $V_{PRG}$. As the strength of $V_{PRG}$ increases, domains switch more deterministically causing asymptotic decay of variation for higher $\mu_{com}$. However, the presence of sharp kinks in the $\sigma_{com}$ profile can be identified for scaled devices. These sharp transitions correspond to the non-uniform coercive field distribution in the FE film, causing abruptly varying gradient in conductance switching, as can be understood from the simulation data provided in Fig. 3(a)-(b). Though the magnitude of variations reduces with increasing $V_{Read}$, their nature remains almost insensitive and is characteristic to the device structure (i.e., dispersion of domains in the FE layer). Hence, proper understanding and extraction of device dependent variation characteristics is extremely important for investigating the impact of device non-idealities at the crossbar array level. 
\begin{figure} [t]
  \begin{center}
  \includegraphics[width=3.4in]{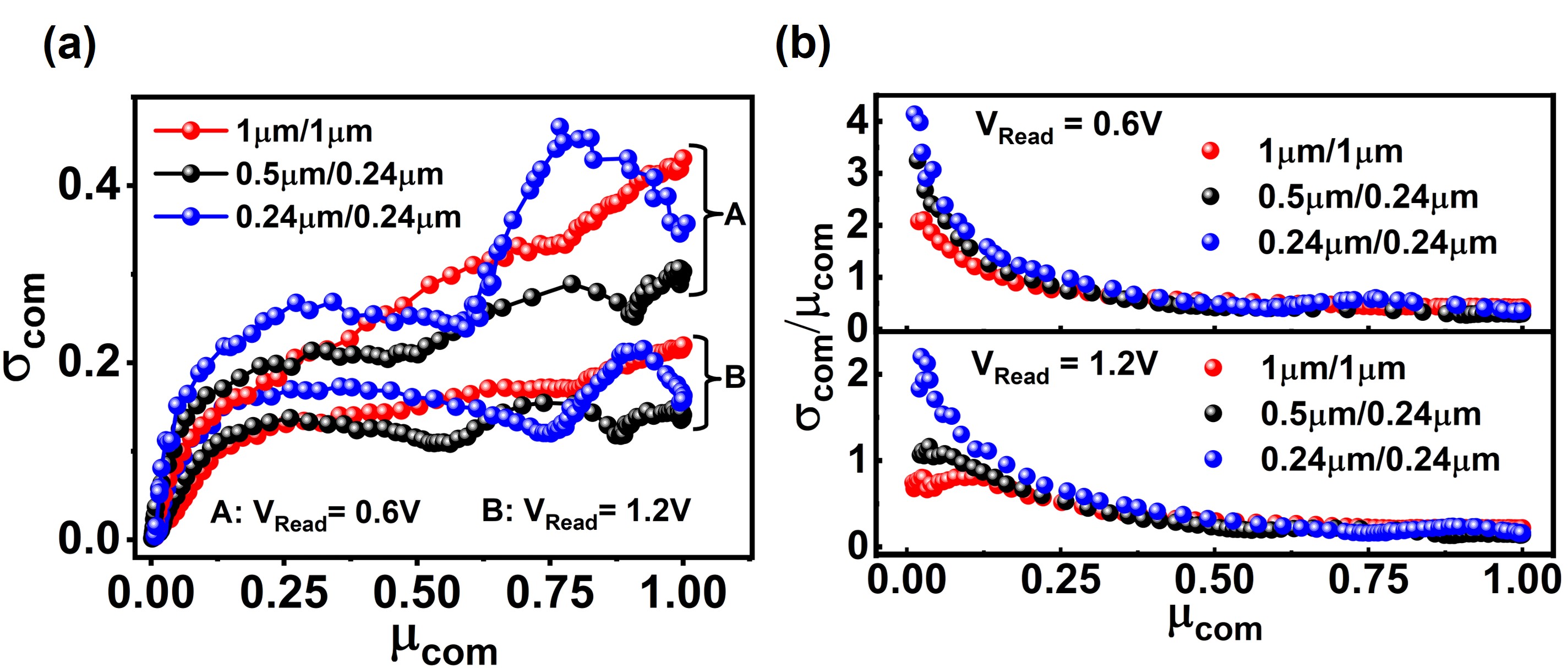}\\
  \caption{(a) The standard deviation, $\sigma_{com}$, as a function of mean, $\mu_{com}$, of the variation in FeFET programming combining both C2C and D2D measurement data. (b) The severity of variations, $\sigma_{com}$/$\mu_{com}$, has been plotted against different values of mean, $\mu_{com}$, at $V_{Read}$ of 0.6V and 1.2V.}
  \label{variation_modeling}
  \end{center}
\end{figure}

Next, we derive an empirical model equation that best fits the variation characteristics shown in Fig. 4(a) employing a higher order polynomial function. The generalized fitted variation equation as a function of the mean programmed conductance state can be expressed as:
\begin{equation}\label{history_parameter}
    \sigma_{com'} = \sum_{i=0}^{n} C_{i}\mu_{com'}^{i}
\end{equation}
where, $\mu_{com'}$ and $\sigma_{com'}$ are the fitted equivalent of $\mu_{com}$ and $\sigma_{com}$ respectively. The coefficients in Eqn. (5) can be derived to minimize the approximation error. For instance, the parameters for 1$\mu$m/1$\mu$m device at $V_{Read}$ of 1.2V have been extracted as: \(C_{0} = 0.0258\), \(C_{1} = 0.788\) \(C_{2} = -0.0214\), and \(C_{3} = 2.1\times 10^{-4}\). The relative variation ($\sigma_{com}$/$\mu_{com}$) plotted in Fig. 4(b) suggests that a significant fraction of the available conductance states undergoes a remarkably high amount of variation and severity increases at lower $V_{Read}$. 

\subsection {Prior Formulation for BNN Training}
Simple prior formulation based on a generic variation model utilized in prior works cannot account for hardware-specific dependencies of the variation effects on device scaling, read noise, presence of sharp transitions in variation spectra of scaled devices, among others. Hence, there is an obvious need for reformulating the prior.   Considering no correlation among the variations of neighboring devices, we reformulate the prior, \(P(w)\), employing a univariate Gaussian distribution. While the mean of the prior, $\mu_{p}$, for each weight follows the mean of the respective posterior, $\mu_{q}$, at any iteration, the broadening parameter, $\sigma_{p}$, is estimated from the relative variation (as provided in Eqn. (5)) experienced by the weight equivalent conductance, $\sigma_{com'}$. Such prior formulation approach enables us to efficiently encode the exact variation structural characteristics in the probability distribution of the network weights. This sharply contrasts prior works on BNN, where a Gaussian model with a fixed amount of variation was considered as the prior for all the weights \cite{b12, b13}.

\section {Performance Evaluation of Proposed Framework}

\begin{figure*} [t] 
  \begin{center}
  \includegraphics[width=5.6in]{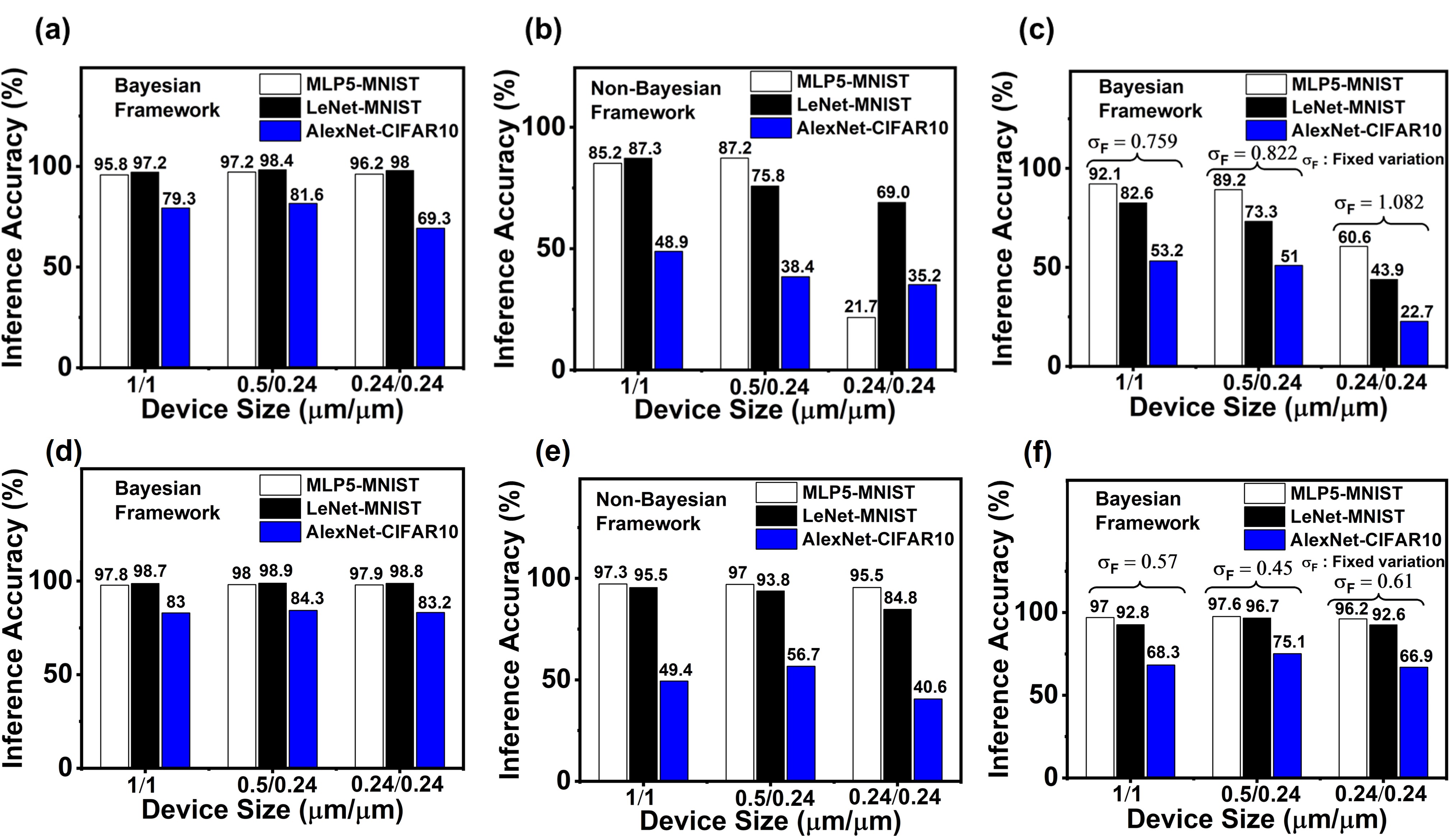}\\
  \caption{Bar-chart comparison of inference accuracy for different network models under variations corresponding to different device sizes (following Eqn. (5)) at $V_{Read}$ of 0.6V, employing (a) proposed Bayesian and (b) Non-Bayesian frameworks. (c) Inference accuracy of network models trained under Bayesian framework but all network weights are subjected to a fixed amount of variation, $\sigma_{F}$, irrespective of programmed conductance state. (d)-(f) Inference performance results for different network models evaluated at $V_{Read}$ of 1.2V applying the same respective schemes as in (a)-(c). The inference outputs in the Bayesian frameworks have been derived by injecting noises (Eqn. (5)) to the trained mean weights and averaging over five runs.}  \label{variation_modeling}
  \end{center}
\end{figure*}

The performance of the proposed framework was evaluated for three different neural network architectures \cite{b27}: five layered MLP5, LeNet, and AlexNet on MNIST \cite{b28} and CIFAR10 \cite{b29} datasets. The algorithm for Bayesian learning was developed following the “Bayes by Backprop” method \cite{b16} and was implemented in PyTorch\footnote{Our implementation is based on a modified version of  an open-source codebase available at \url{https://github.com/kumar-shridhar/PyTorch-BayesianCNN}.}. The local re-parameterization trick has been employed to reduce computational overhead by translating expensive sampling operation over noisy distributions from high-dimensional weight space to the lower dimensional activation level \cite{b30}. The layer-wise KL loss between prior and posterior (\(P(w)\sim N(\mu_{p}, \sigma_{p}); q(w)\sim N(\mu_{q}, \sigma_{q})\)) has been calculated following \cite{b31}:  \(KL(P(w)||q(w))= log(\sigma_{q}/\sigma_{p}) + (\sigma_{p}^2/2\sigma_{q}^2) + (\mu_{p}-\mu_{q})^2/2\sigma_{q}^2 - 1/2\). 10\% of KL loss was added to the standard likelihood loss to derive total loss of the network at every iteration. The network has been trained using Adam optimizer with an efficient learning-rate scheduler and input batch size of 128. The ideal software-based inference accuracies of MLP5-MNIST, LeNet-MNIST, and AlexNet-CIFAR10 (architecture-dataset format) without any variations are 98$\%$, 99.1$\%$, and 85.4$\%$, respectively. The robustness of the Bayesian framework was assessed by comparing the inference performance of respective network models with that of the standard non-Bayesian equivalent, where networks are trained iteratively under variations injected into the weights following the canonical weight variation model\cite{b11}. Fig. 5(a)-(b) and (d)-(e) demonstrate the comparative inference results as bar chart representations for Bayesian and non-Bayesian counterparts under variations corresponding to different device sizes at $V_{Read}$ of 0.6V and 1.2V, respectively. The robustness was evaluated under larger variations observed at a lower $V_{Read}$ of 0.6V, where non-Bayesian trained networks suffer from substantial accuracy loss, which becomes more severe as the deviation increases with device down-scaling. The more considerable accuracy degradation in AlexNet is primarily due to its deeper and more complex network architecture where variation across weights in all the layers gets accumulated to cause more ambiguity in the inference output. The proposed Bayesian framework dramatically minimizes the accuracy loss by retaining the near-ideal baseline accuracies %(maximum loss $\sim2\%$)
for two shallow networks (MLP5 and LeNet) and exhibiting minimal accuracy drop for AlexNet. The accuracy loss for AlexNet architecture on CIFAR10 dataset has been observed to be 6.1$\%$, 3.8$\%$, and 16.1$\%$ with respect to the ideal accuracy value for 1$\mu$m/1$\mu$m, 0.5$\mu$m/0.24$\mu$m, and 0.24$\mu$m/0.24$\mu$m sized devices respectively, at $V_{Read}$ of 0.6V. To make our study more meaningful, inference comparison results are also provided for a $V_{Read}$ of 1.2V, as this voltage point is found to be optimal with respect to accuracy, training convergence, energy consumption, and conductance bit precision (as mentioned in the subsequent discussion). As expected, the smaller conductance variations for all device sizes at higher read voltages yield more improved accuracies. The result underscores the usefulness of our framework to provide robust and efficient inference under variations imposed even by highly scaled devices at smaller read voltages. 

The benefits of adopting device-specific entire variation spectra as prior for the BNN (including interplay with device size and read voltage) instead of employing a uniform and fixed variation model for all network weights \cite{b12, b13} is substantiated by the accuracy comparison results provided in Fig. 5(a) and (c), and Fig. 5(d) and (f).  %for different network models. % 
The single variation value, $\sigma_{F}$, used for each device size, as mentioned in Fig. 5(c) and (f), was estimated by averaging conductance variations over the entire operating range of the device (see Fig. 4(b)). %While both approaches yield similar performances for shallow networks by providing close to ideal accuracies, our proposed scheme outperforms the uniform variation based method specifically for the deeper network exposed to high level of weight fluctuations.
Our proposed scheme has been found to outperform the uniform variation based method in terms of accuracy for all three network architectures exposed to different degree of weight fluctuations corresponding to different device sizes. 
It offers a noteworthy improvement in accuracy by 46.6$\%$, 54.1$\%$, and 35.2$\%$ for AlexNet-CIFAR10, LeNet-MNIST, and MLP5-MNIST respectively for the smallest device size operating at the lowest $V_{Read}$ - thereby substantiating the need for such hardware-software co-design efforts from a scalability perspective on complex machine learning tasks.
\begin{figure} [t]
  \begin{center}
  \includegraphics[width=2.5in]{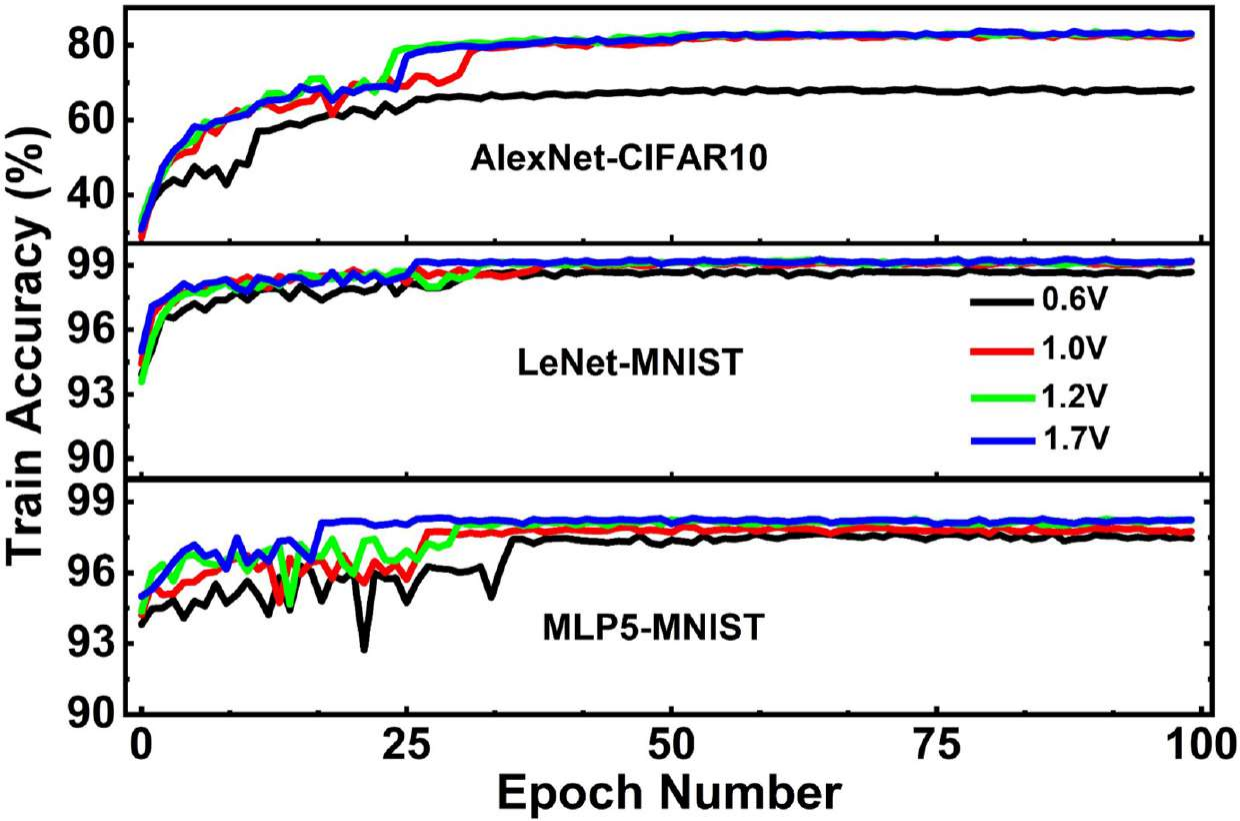}\\
  \caption{Training dynamics of the proposed approach under variations corresponding to 0.24$\mu$m/0.24$\mu$m device at different read voltages.}
  \label{Training_dynamics}
  \end{center}
\end{figure}

The impact of read voltage on the training dynamics of the proposed approach was also investigated for  variations corresponding to the smallest device size. As revealed in Fig. 6, the higher read voltage not only improves the accuracy by offering lower conductance fluctuations but also accomplishes a stable training convergence at a relatively smaller number of epochs. However, the higher read operation causes more power consumption and limits the available number of conductance states in the operating range. Thus, a $V_{Read}$ of around 1.2V could be an optimal solution to provide a reasonable trade-off between accuracy and power consumption.

%The present variation-tolerant design is based on the true variation model derived from straightforward experimental data and is found to offer superior performance in terms of training dynamics and accuracy over some recent works on Bayesian learning where variations are based on a hypothetical model \cite{b12, b13}. Our model defines collective non-idealities at individual device level, including read and write disturbances. Other process-variation-related issues and the spatially correlated noises among neighboring FeFET at the crossbar array might be considered in our future work to examine effectiveness of the framework under complete hardware-level variations.

\section{Conclusion}
In summary, we propose a novel device-algorithm co-design approach for reliable ferroelectric in-memory computing where a comprehensive conductance variation model derived by systematically characterizing FeFET devices is coupled to Bayesian learning based uncertainty optimizer to alleviate the impact of device level non-idealities. Incorporating dependencies of variation properties with operating voltage conditions and device size during the training process is shown to play a significant role in minimizing accuracy loss for complex datasets and deeper networks. The main advantage of this approach against hardware-in-the-loop training is that this will be a one-time training process without costly iterative training. Other process-variation related issues like spatially correlated noise effects among neighboring FeFET devices in the crossbar array can be considered in  future work to extend the efficacy of the proposed framework.

\section*{Acknowledgment}
The authors would like to acknowledge GlobalFoundries, Dresden, Germany, for providing FeFET testing devices. We are thankful to Suma George Cardwell and Cale Douglas Crowder from Sandia National Labs for their valuable suggestions regarding the work.
This material is based upon work supported primarily by the National Science Foundation under Grant No. CNS 2137259 - Center for Advanced Electronics through Machine Learning (CAEML) and its industry members. The characterization of FeFET device non-idealities is supported by the U.S.
Department of Energy, Office of Science, Office of Basic Energy Sciences Energy Frontier Research Centers program under Award No. DE-SC0021118 and the National Science Foundation under Grant No.  2347024. 

%\vspace{-2mm}

%\bibliography{bare_conf}
%\bibliographystyle{IEEEtran}


\end{document}